# F-theory superspace backgrounds

Warren Siegel and Di Wang

*CNYITP*
*Stony Brook University*

**Abstract**

F-theory is the theory proposed to incorporate superstring theory in a way such that STU dualities are manifest. A useful description uses a current superalgebra on a higher-dimensional worldvolume, following from an action for a selfdual gauge field.

Here the group "metric" appearing in the Schwinger (central charge) term of this current superalgebra is generalized to a tensor, in analogy to the usual generalization of the structure constants to the torsion (and curvature). This allows introduction of a massless background describing F-supergravity on the original bosonic worldvolume. The isotropy group is represented on superspace, while the (exceptional) symmetry is represented on the worldvolume.

As an example, we solve off shell the linearized superspace constraints of the massless sector of the F-theory that generalizes the N=2 supergravity (+ matter) of 3D S(tring)-theory, the corresponding manifestly T-dual theory of T-theory, and the N=1 supergravity of 4D M-theory. The results for the prepotential, its gauge transformation, and action agree with those that were derived previously without reference to the current algebra of the full F-theory.

# Contents

# 1 Introduction

## 1.1 History

T-theory [1, 2] was originally developed from current algebra to describe the massless gauge fields (not compactification scalars) of strings and superstrings as coset spaces G/H, with G an orthosymplectic group (dictated by the current algebra) and H the product of left and right Lorentz groups (required to be independent by T-duality). The current superalgebra defines the torsion in superspace, whose constraints yield the background field equations. A concept of sectioning was implied by (the $\sigma$ part of) the Virasoro constraints. A metric on the tangent space was considered, but it was constrained to be pure gauge (as in Cartan's description of curved space).

F-supergravity was a generalization of this massless field theory that incorporated the exceptional symmetries of extended supergravities as G, generalizing T-theory sectioning so that gauge transformations preserved the fields' membership in the group [3, 4]. In the case of the 3D superstring, an off-shell superspace formulation was found (at the linearized level) in terms of a prepotential $H^{\alpha}$ and gauge parameter $\lambda^{\alpha}$ [5].

An F-theory current algebra was found that derived this coset structure and sectioning from a generalization of string theory defined on a higher-dimensional worldvolume [6]. In the original example of the 3D string, spacetime $X$ appeared as a (6-dimensional-)worldvolume 2-form gauge field $X_{mn}$ , with selfdual field-strength $P_{mnp}$. Dropping worldvolume "time" $\tau$ (as coordinate and index), that became the current on 5-dimensional "space" $\sigma$,

$$P_m = P^{mn} = \frac{1}{6}\varepsilon^{mnpqr}P_{pqr}$$

The relation to other formulations of string theory involved not only Virasoro-based sectioning but also the Gauss constraint of this gauge field:

| | worldvolume | Gauss ($\mathcal{U}$) | worldsheet |
|---|---|---|---|
| sectioning | F-theory | → | T-theory |
| section ($\mathcal{S}$) | ↓ | | ↓ |
| no sectioning | M-theory | → | S-theory |

Generalization to other dimensions and superspace soon followed [7, 8], with superalgebras for "flat"-space (vacuum) currents $\mathring{\triangleright}$ [8]:

$$i[\mathring{\triangleright}_M(1), \mathring{\triangleright}_N(2)\} = f_{MN}{}^P\mathring{\triangleright}_P\delta(1-2) + \eta_{MN}{}^p(\mathring{\mathcal{D}}_p(2) - \mathring{\mathcal{D}}_p(1))\delta(1-2)$$

in terms of structure constants $f$ and a generalized group "metric" $\eta$. The minimal set of supercurrents for the supersymmetric case, as for S(tring)-theory [9] and T-theory, are $DP\Omega$ (where $\Omega$ is "dual" to the usual spinor $D$), with nonvanishing constant tensors

$$f_{DD}{}^P,\ f_{DP}{}^\Omega; \quad \eta_{D\Omega},\ \eta_{PP}$$

However, for first-quantized treatment of Lorentz-covariant derivatives and currents the spin currents $S$ and their dual $\Sigma$ are also useful [10, 11].

In [12] it was found that the inclusion of the isotropy group currents $S_{mn}$ required a non-constant vielbein $\mathring{\mathcal{E}}_{a}{}^{m}$ (in $\mathring{\mathcal{D}}_{a} = \mathring{\mathcal{E}}_{a}{}^{m}\partial_{m}$), depending on the isotropy coordinates, for the flat worldvolume $\sigma^{m}$ to preserve the invariance of the group metric $\eta_{mn}{}^{p}$. In [13] these results were extended from the isotropy group to the current-algebra group (defined by the structure constants $f_{MN}{}^{P}$), requiring a supersymmetric extension of the worldvolume, and the replacement of the dual current $\Omega^{\mu}_{m}$ with the simpler $\Omega^{\mu}$ of S-theory and T-theory. This apparently resolved a problem with the worldvolume (generalized Virasoro $\mathcal{S}_{m}$) transformation of the fermionic current $D_{\mu}$, and perhaps also with the consistent introduction of a massless supersymmetric background $E_{A}{}^{M}$.

## 1.2 Present

Unfortunately, this generalization led to new complications (which will not be discussed here). In the meantime, it was discovered that the problem with $D_{\mu}$ could be solved in the same way as it had already been solved for the usual bosonic current $P_{m}$, by applying the Gauss constraint $\mathcal{U}$. (See details below.) As a result, the dual $\eta^{MN}{}_{p}$ of the group metric of the current algebra, which appears in the Virasoro operators, need not be invariant under that group.

Further constraints are dropped by recognizing that, unlike T-theory or bosonic F-theory, the metric cannot always be set to its flat value $\eta$. Thus, when introducing a background for the currents as

$$\triangleright_{A} \equiv E_{A}{}^{M}\mathring{\triangleright}_{M}\,, \quad \mathcal{D}_{a} \equiv \mathcal{E}_{a}{}^{m}\mathring{\mathcal{D}}_{m}$$
$$\Rightarrow \quad i[\triangleright_{A}, \triangleright_{B}\} = T_{AB}{}^{C}\triangleright_{C}\delta + (G_{AB}{}^{c}\mathcal{D}_{c}((2)-(1))\delta$$

we have a "curved" metric (with flat indices)

$$G_{AB}{}^{c} \equiv E_{A}{}^{M}E_{B}{}^{N}\eta_{MN}{}^{p}\mathcal{E}_{p}{}^{c}$$

The lower-dimension components of $G$ take their vacuum values $\eta$, while some higher-dimension components become field strengths, analogously to the torsion $T_{AB}{}^{C}$. Consequently, the vielbein $E$ does not satisfy a complete "orthogonality" condition as it does in the free case (or as in T-theory, where the background vielbein is an element of an orthosymplectic group in an appropriate gauge).

However, since $G_{PP}{}^{c} = \eta_{PP}{}^{c}$, $\mathcal{E}_{a}{}^{m}$ is still an element of the coset G/H (in an appropriate gauge), where H is the isotropy group and G is the exceptional symmetry group. Thus the gauge group H acts on the index $a$ as on all spacetime indices, while G acts only on the curved index $m$: H transforms all the spacetime indices of superspace (like the local Lorentz group of gravity), while G acts on the worldvolume. (I.e., $\sigma^{m}$ is a representation of G. This differs from the interpretation of [14], where G was taken to act on compactified spacetime dimensions, but whose partial derivatives were also interpreted as “central charges”.) G is then the subgroup of linear coordinate transformations GL(d) on the d-dimensional worldvolume that leaves the flat metric $\eta_{mn}{}^{\mathscr{p}}$ invariant. Since the exceptional groups have no graded extensions in general, this requires the worldvolume to be bosonic.

# 2 Background

## 2.1 Vielbein, Torsion and Gauge Transformation

The expression for the torsion follows from its definition in the current algebra (with background):

$$T_{AB}{}^{C} = E_A{}^M E_B{}^N f_{MN}{}^P E_P{}^C + (\delta^E_{[A|}\delta^C_D - \tfrac{1}{2} G_{[A|D}{}^{\mathscr{f}} G^{CE}{}_{\mathscr{f}})(\nabla_E E_{|B)}{}^M) E_M{}^D$$

$$\nabla_A \equiv E_A{}^M \mathring{\nabla}_M$$

We have used, for any superspace function $f$,

$$i[\rhd_A, f\} = (\nabla_A f)\delta\,, \quad \mathcal{D}_{\mathscr{a}} f = \tfrac{1}{2} G^{BC}{}_{\mathscr{a}} (\nabla_B f)\rhd_C\,, \quad \nabla_\Omega = 0$$

We define also the “dual” metric

$$G^{AB}{}_{\mathscr{c}} \equiv E_M{}^A E_N{}^B \eta^{MN}{}_{\mathscr{p}} \mathcal{E}_{\mathscr{c}}{}^{\mathscr{p}}$$

The Bianchi identities are

$$T_{[AB}{}^E T_{C)E}{}^D + (\delta^E_{[A}\delta^D_F - \tfrac{1}{2} G_{F[A|}{}^{\mathscr{g}} G^{DE}{}_{\mathscr{g}})\nabla_E T_{|BC)}{}^F = 0$$

$$T_{C(A}{}^E G_{B]E}{}^{\mathscr{d}} = F_{ABC}{}^{\mathscr{d}} - \tfrac{1}{2} F_{C(AB]}{}^{\mathscr{d}} \quad \Leftrightarrow \quad T_{AB}{}^E G_{CE}{}^{\mathscr{d}} - \tfrac{1}{2} T_{C[A}{}^E G_{B)E}{}^{\mathscr{d}} = -\tfrac{3}{4} F_{C[AB)}{}^{\mathscr{d}}$$

where

$$F_{ABC}{}^{\mathscr{d}} \equiv (\nabla_C G_{AB}{}^m)\mathcal{E}_m{}^{\mathscr{d}} = \nabla_C G_{AB}{}^{\mathscr{d}} + \mathcal{T}_{C\mathscr{e}}{}^{\mathscr{d}} G_{AB}{}^{\mathscr{e}}$$

$$\mathcal{T}_{A\mathscr{b}}{}^{\mathscr{c}} \equiv (\nabla_A \mathcal{E}_{\mathscr{b}}{}^m)\mathcal{E}_m{}^{\mathscr{c}}$$

$\nabla G$ and $\mathcal{T}$ appear in the Bianchi identities only in this combination because $G_{AB}{}^{\mathscr{c}}\mathcal{E}_{\mathscr{c}}{}^m = G_{AB}{}^m = EE\eta$ is what actually appears in $[\rhd, \rhd\}$. The only pieces of $F$ and $\mathcal{T}$ that are covariant are those that appear in the Bianchi identities.

Gauge transformations are generated by

$$\delta = i[\textstyle\int d\sigma\, \Lambda^A \rhd_A, \quad ]$$

Acting on another $\triangleright$, the former gives

$$(\delta E_A{}^M)E_M{}^B = \Lambda^C T_{CA}{}^B - (\delta_A^D \delta_C^B - G_{AC}{}^{\mathscr{e}} G^{BD}{}_{\mathscr{e}})\nabla_D \Lambda^C$$

We have chosen $\mathcal{S}$ terms in $\delta\triangleright$ so that $\delta E$ is orthogonal, even though $E$ isn't.

Due to the worldvolume gauge symmetry, the algebra also needs to satisfy a $\mathcal{U}$ constraint, with:

$$\mathcal{U}_{mn} = U^{n\mathscr{n}}_{m\mathscr{m}} P_n \mathcal{D}_{\mathscr{n}} \approx 0\,, \quad U^{n\mathscr{n}}_{m\mathscr{m}} \equiv \delta^n_m \delta^{\mathscr{n}}_{\mathscr{m}} - \eta_{pm}{}^{\mathscr{n}} \eta^{pn}{}_{\mathscr{m}} = U^{\varpi}_{m\mathscr{m}} U^{n\mathscr{n}}_{\varpi}$$

Here $U^{\varpi}_{m\mathscr{m}}$ stands for the decomposition of $U^{n\mathscr{n}}_{m\mathscr{m}}$ with respect to the representation of worldvolume gauge symmetry. For example in this 3D case, we have $\varpi$ living in a 5:

$$U^{n\mathscr{n}}_{m\mathscr{m}} = \delta^{n}_{[n_1} \delta^{[m_1}_{n_2]} \delta^{m_2]}_{\mathscr{m}} = \delta^{n}_{[n_1} \delta^{\varpi}_{n_2]} \delta^{[m_1}_{\varpi} \delta^{m_2]}_{\mathscr{m}}$$

There is also a $\mathcal{V}$ constraint, acting on worldvolume differentials:

$$\mathcal{V}_{\varpi m} = U^{n\mathscr{n}}_{\varpi} \eta_{nm}{}^{\mathscr{m}} \mathcal{D}_{\mathscr{m}} \mathcal{D}_{\mathscr{n}} \approx 0$$

We would use:

$$\mathcal{U}_{mn} f = U^{n\mathscr{n}}_{m\mathscr{m}} \mathcal{D}_{\mathscr{n}} \nabla_n f$$
$$\mathcal{U}_{mn}(f, g) = U^{n\mathscr{n}}_{m\mathscr{m}} (\nabla_n f)(\mathcal{D}_{\mathscr{n}} g)$$

for the two derivatives in $\mathcal{U}$ acting on the same function $f$ or different functions $f$, g. We also define the action of $\mathcal{U}_{\varpi}$ and $\mathcal{V}_{\varpi m}$ on $f$ or $f$, $g$ in the same way.

$G_{AB}{}^{\mathscr{c}}$ is covariant only up to $\mathcal{U}$ terms:

$$(\delta G_{AB}{}^{\mathscr{a}})\mathcal{D}_{\mathscr{a}} \sim \mathcal{U}$$

In components, we have:

$$\begin{aligned}
&\delta G_{DD}{}^{\mathscr{a}} \mathcal{D}_{\mathscr{a}} = 0 \\
&\delta G_{DP}{}^{\mathscr{a}} \mathcal{D}_{\mathscr{a}} = \delta G_{\alpha a}{}^{\mathscr{a}} \mathcal{D}_{\mathscr{a}} = U^{b\mathscr{a}}_{a\mathscr{b}} \nabla_b \Lambda_\alpha{}^{\mathscr{b}} \mathcal{D}_{\mathscr{a}} = \mathcal{U}_{a\mathscr{b}}(\Lambda_\alpha{}^{\mathscr{b}},) \\
&\delta G_{PP}{}^{\mathscr{a}} \mathcal{D}_{\mathscr{a}} = 0 \\
&\delta G_{D\Omega}{}^{\mathscr{a}} \mathcal{D}_{\mathscr{a}} = \delta G_\alpha{}^\beta{}_{\mathscr{b}}{}^{\mathscr{a}} \mathcal{D}_{\mathscr{a}} = \delta^\beta_\alpha U^{b\mathscr{a}}_{a\,\mathscr{b}} \nabla_b \Lambda^a \mathcal{D}_{\mathscr{a}} = \delta^\beta_\alpha \mathcal{U}_{b\mathscr{b}}(\Lambda^b,) \\
&\delta G_{P\Omega}{}^{\mathscr{a}} \mathcal{D}_{\mathscr{a}} = \delta G_{a\mathscr{b}}{}^{\alpha\mathscr{a}} \mathcal{D}_{\mathscr{a}} = U^{b\mathscr{a}}_{a\mathscr{b}} \nabla_b \Lambda^\alpha \mathcal{D}_{\mathscr{a}} = \mathcal{U}_{a\mathscr{b}}(\Lambda^\alpha,) \\
&\delta G_{\Omega\Omega}{}^{\mathscr{a}} \mathcal{D}_{\mathscr{a}} = 0
\end{aligned}$$

Torsions are also covariant only up to $\mathcal{U}$ terms, which can be dropped off when coupled with $\rhd_\Omega (= \mathcal{D}\theta)$ [8, 15] or $\rhd_P$. Here we list torsions with extra pieces up to dimension 1:

$$
\begin{aligned}
\delta T_{DD}{}^{\Omega} \rhd_\Omega &= \delta T_{\alpha\beta\gamma}{}^{\mathscr{a}} \mathcal{D}_{\mathscr{a}} \theta^\gamma = -U^{c\mathscr{a}}_{b\mathscr{b}} \nabla_c \Lambda_{(\alpha}{}^{\mathscr{b}} \gamma^b_{\beta)\gamma} \mathcal{D}_{\mathscr{a}} \theta^\gamma \\
&= -\gamma^b_{\gamma(\beta} \mathcal{U}_{b\mathscr{b}} (\Lambda_{\alpha)}{}^{\mathscr{b}}, \theta^\gamma) \\
\delta T_{DP}{}^{\Omega} \rhd_\Omega &= \delta T_{\alpha a \beta}{}^{\mathscr{a}} \mathcal{D}_{\mathscr{a}} \theta^\beta = U^{c\mathscr{a}}_{a\mathscr{b}} \mathcal{D}_{\mathscr{a}} \theta^\beta \nabla_c (\Lambda^d \eta_{de}{}^{\mathscr{b}} \gamma^e_{\alpha\beta} + \nabla_{(\alpha} \Lambda_{\beta)}{}^{\mathscr{b}}) \\
&= \mathcal{U}_{a\mathscr{b}} (\gamma^e_{\alpha\beta} \eta_{de}{}^{\mathscr{b}} \Lambda^d + \nabla_{(\alpha} \Lambda_{\beta)}{}^{\mathscr{b}}, \theta^\beta) \\
\delta T_{DP}{}^{P} \rhd_P &= \delta T_{\alpha a}{}^{b} \rhd_b = \eta^{bc}{}_{\mathscr{a}} U^{d\mathscr{a}}_{a\mathscr{b}} \nabla_c \nabla_d \Lambda_\alpha{}^{\mathscr{b}} \rhd_b = U^{d\mathscr{a}}_{a\mathscr{b}} \mathcal{D}_{\mathscr{a}} \nabla_d \Lambda_\alpha{}^{\mathscr{b}} \\
&= \mathcal{U}_{a\mathscr{b}} \Lambda_\alpha{}^{\mathscr{b}} \\
\delta T_{PP}{}^{\Omega} \rhd_\Omega &= \delta T_{ab}{}^{\mathscr{a}}{}_{\alpha} \mathcal{D}_{\mathscr{a}} \theta^\alpha = -\nabla_\alpha \eta_{c[a}{}^{\mathscr{b}} U^{d\mathscr{a}}_{b]\mathscr{b}} \mathcal{D}_{\mathscr{a}} \theta^\alpha \nabla_d \Lambda^c \\
&= -\eta_{c[a}{}^{\mathscr{b}} \mathcal{U}_{b]\mathscr{b}} (\nabla_\alpha \Lambda^c, \theta^\alpha) \\
\delta T_{D\Omega}{}^{\Omega} \rhd_\Omega &= \delta T_{\alpha\mathscr{a}}{}^{\beta\mathscr{b}}{}_{\gamma} \mathcal{D}_{\mathscr{b}} \theta^\gamma = -\gamma^a_{\alpha\gamma} U^{b\mathscr{b}}_{a\mathscr{a}} \nabla_b \Lambda^\beta \mathcal{D}_{\mathscr{b}} \theta^\gamma \\
&= -\gamma^a_{\alpha\gamma} \mathcal{U}_{a\mathscr{a}} (\Lambda^\beta, \theta^\gamma) \\
\delta T_{PP}{}^{P} \rhd_P &= \delta T_{ab}{}^{c} \rhd_c = \eta_{d[a}{}^{\mathscr{b}} \eta^{fc}{}_{\mathscr{a}} U^{e\mathscr{a}}_{b]\mathscr{b}} \nabla_e \nabla_f \Lambda^d \rhd_c = \eta_{d[a}{}^{\mathscr{b}} U^{e\mathscr{a}}_{b]\mathscr{b}} \mathcal{D}_{\mathscr{a}} \nabla_e \Lambda^d \\
&= \eta_{d[a}{}^{\mathscr{b}} \mathcal{U}_{b]\mathscr{b}} \Lambda^d \\
\delta T_{P\Omega}{}^{\Omega} \rhd_\Omega &= \delta T_{a\mathscr{a}}{}^{\alpha\mathscr{b}}{}_{\beta} \mathcal{D}_{\mathscr{b}} \theta^\beta = \frac{1}{2} U^{b\mathscr{b}}_{a\mathscr{a}} \nabla_\beta \nabla_b \Lambda^\alpha \mathcal{D}_{\mathscr{b}} \theta^\beta \\
&= \frac{1}{2} \mathcal{U}_{a\mathscr{a}} (\nabla_\beta \Lambda^\alpha, \theta^\beta)
\end{aligned}
$$

When comupting $\delta T_{DP}{}^{P}$ or $\delta T_{PP}{}^{P}$, we can convert a spacetime derivative, together with a current, into a worldvolume derivative:

$$
\eta^{cd}{}_{\mathscr{a}} \nabla_d \Lambda \rhd_c = 2[\frac{1}{2} \eta^{cd}{}_{\mathscr{a}} \rhd_c \rhd_d, \Lambda] = 2\mathcal{D}_{\mathscr{a}} \Lambda
$$

## 2.2 Gauge transformations of the gauge parameters

Gauge transformations of the gauge parameters are defined as operations on spacetime gauge transformation parameters $\Lambda$ leaving the whole generator $\delta = i[\int d\sigma \Lambda^A \rhd_A, \quad ]$ invariant mod $\mathcal{S}$ or $\mathcal{U}$ constraints. Right now we have various kinds of them:

- Up to $\mathcal{S}$ sectioning, which generates a worldvolume diffeomorphism:

$$
\begin{aligned}
\delta\Lambda^{A}\triangleright_{A} &= -\eta^{BA}{}_{\mathscr{a}}\nabla_{B}\lambda^{\mathscr{a}}\triangleright_{A} = -2\mathcal{D}_{\mathscr{a}}\lambda^{\mathscr{a}} \\
\delta H_{D}{}^{\Omega}\triangleright_{\Omega} &= \delta H_{\alpha\beta}{}^{\mathscr{a}}\mathcal{D}_{\mathscr{a}}\theta^{\beta} = \gamma^{a}_{\alpha\beta}U^{b\mathscr{a}}_{a\mathscr{b}}\nabla_{b}\Lambda^{\mathscr{b}}\mathcal{D}_{\mathscr{a}}\theta^{\beta} \\
&= \gamma^{a}_{\alpha\beta}\mathcal{U}_{a\mathscr{b}}(\lambda^{\mathscr{b}},\theta^{\beta}) \\
\delta H_{P}{}^{\Omega}\triangleright_{\Omega} &= \delta H_{a\alpha}{}^{\mathscr{a}}\mathcal{D}_{\mathscr{a}}\theta^{\alpha} = U^{b\mathscr{a}}_{a\mathscr{b}}\nabla_{\alpha}\nabla_{b}\Lambda^{\mathscr{b}}\mathcal{D}_{\mathscr{a}}\theta^{\alpha} \\
&= \mathcal{U}_{a\mathscr{b}}(\nabla_{\alpha}\lambda^{\mathscr{b}},\theta^{\alpha}) \\
\delta H_{P}{}^{P}\triangleright_{P} &= \delta H_{a}{}^{b}\triangleright_{b} = U^{e\mathscr{a}}_{a\mathscr{b}}\eta^{bd}{}_{\mathscr{a}}\nabla_{d}\nabla_{e}\Lambda^{\mathscr{b}}\triangleright_{b} \\
&= \mathcal{U}_{a\mathscr{b}}\lambda^{\mathscr{b}} \\
\delta(others) &= 0
\end{aligned}
$$

  All extra terms are proportional to $\mathcal{U}$.

- Up to $\mathcal{U}$ sectioning on $\theta$:

$$
\begin{aligned}
\delta\Lambda^{\Omega}\triangleright_{\Omega} &= \delta\Lambda^{\mathscr{a}}{}_{\alpha}\mathcal{D}_{\mathscr{a}}\theta^{\alpha} = -U^{a\mathscr{a}}_{\varpi}\nabla_{a}\Lambda^{\varpi}_{\alpha}\mathcal{D}_{\mathscr{a}}\theta^{\alpha} \\
&= -\mathcal{U}_{\varpi}(\Lambda^{\varpi}_{\alpha},\theta^{\alpha}) \\
\delta H_{D}{}^{\Omega}\triangleright_{\Omega} &= \delta H_{\alpha\beta}{}^{\mathscr{a}}\mathcal{D}_{\mathscr{a}}\theta^{\beta} = U^{a\mathscr{a}}_{\varpi}\nabla_{a}\nabla_{(\beta}\Lambda^{\varpi}_{\alpha)}\mathcal{D}_{\mathscr{a}}\theta^{\beta} \\
&= \mathcal{U}_{\varpi}(\nabla_{(\beta}\Lambda^{\varpi}_{\alpha)},\theta^{\beta}) \\
\delta H_{D}{}^{P}\triangleright_{P} &= \delta H_{\alpha}{}^{a}\triangleright_{a} = -\eta^{ab}{}_{\mathscr{a}}\nabla_{b}U^{c\mathscr{a}}_{\varpi}\nabla_{c}\Lambda^{\varpi}_{\alpha}\triangleright_{a} \\
&= -\mathcal{U}_{\varpi}\Lambda^{\varpi}_{\alpha} \\
\delta H_{P}{}^{\Omega}\triangleright_{\Omega} &= \delta H_{a\alpha}{}^{\mathscr{a}}\mathcal{D}_{\mathscr{a}}\theta^{\alpha} = \nabla_{a}U^{b\mathscr{a}}_{\varpi}\nabla_{b}\Lambda^{\varpi}_{\alpha}\mathcal{D}_{\mathscr{a}}\theta^{\alpha} \\
&= \mathcal{U}_{\varpi}(\nabla_{a}\Lambda^{\varpi}_{\alpha},\theta^{\alpha}) \\
\delta(others) &= 0
\end{aligned}
$$

  Again, all extra terms are proportional to $\mathcal{U}$.

- Up to $\mathcal{U}$ sectioning on $X$, which generates a worldvolume gauge transformation of $X$:

$$
\begin{aligned}
\delta\Lambda^P \triangleright_P &= \delta\Lambda^a \triangleright_a = -U^{a\mathscr{a}}_{\varpi}\mathcal{D}_{\mathscr{a}}\Lambda^{\varpi}\triangleright_a \\
&= -\mathcal{U}_{\varpi}(\quad , \Lambda^{\varpi}) \\
\delta H_D{}^P \triangleright_P &= \delta H_\alpha{}^a \triangleright_a = \nabla_\alpha U^{a\mathscr{a}}_{\varpi}\mathcal{D}_{\mathscr{a}}\Lambda^{\varpi}\triangleright_a \\
&= \mathcal{U}_{\varpi}(\quad , \nabla_\alpha\Lambda^{\varpi}) \\
\delta H_P{}^\Omega \triangleright_\Omega &= \delta H_{b\alpha}{}^{\mathscr{b}}\mathcal{D}_{\mathscr{b}}\theta^\alpha = -\eta_{ab}{}^{\mathscr{b}}\nabla_\alpha U^{a\mathscr{a}}_{\varpi}\mathcal{D}_{\mathscr{a}}\Lambda^{\varpi}\mathcal{D}_{\mathscr{b}}\theta^\alpha \\
&= -\mathcal{V}_{b\varpi}(\nabla_\alpha\Lambda^{\varpi}, \theta^\alpha) \\
\delta H_P{}^P \triangleright_P &= \delta H_a{}^b \triangleright_b = \nabla_a U^{b\mathscr{a}}_{\varpi}\mathcal{D}_{\mathscr{a}}\Lambda^{\varpi}\triangleright_b - \eta_{ac}{}^{\mathscr{b}}\eta^{bd}{}_{\mathscr{b}}\nabla_d U^{c\mathscr{a}}_{\varpi}\mathcal{D}_{\mathscr{a}}\Lambda^{\varpi}\triangleright_b \\
&= \mathcal{U}_{\varpi}(\quad , \nabla_a\Lambda^{\varpi}) - \mathcal{V}_{a\varpi}\Lambda^{\varpi} \\
\delta(others) &= 0
\end{aligned}
$$

This time, extra terms are proportional to $\mathcal{U}$, $\mathcal{V}$.

# 3 Prepotential

As an explicit example of this superspace background we examine, at the linearized level but off shell, the "minimal" supersymmetric case for the 3D (Type II) superstring. (The corresponding T-theory was analyzed this way in [16], where the existence of this formulation of F-theory was first conjectured.)

The procedure to solve the constraints (as in general gauge theories in superspace) is to work by engineering dimension from the bottom up. The dimension is determined by subtracting that associated with upper indices from that for lower indices, where $[(D, P, \Omega, \mathscr{a})] = (1/2, 1, 3/2, 2)$. The relevant gauge parameters $\Lambda^A$, linearized potentials $H_A{}^B$ ($E_A{}^M = \delta_A^M + H_A{}^M$), metrics $G_{AB}{}^{\mathscr{c}}$, and torsions $T_{AB}{}^C$ are labeled by the indices

| dim | $\Lambda$ | $H\,(\mathcal{H})$ | $G$ | $T$ |
|---|---|---|---|---|
| $-3/2$ | $\Omega$ | | | |
| $-1$ | $P$ | ${}_D{}^\Omega$ | $DD$ | |
| $-1/2$ | $D$ | ${}_D{}^P, {}_P{}^\Omega$ | $DP$ | ${}_{DD}{}^\Omega$ |
| $0$ | $S$ | ${}_D{}^D, {}_P{}^P, {}_\Omega{}^\Omega$ $({}_{\mathscr{a}}{}^{\mathscr{b}})$ | $D\Omega, PP$ | ${}_{DD}{}^P, {}_{DP}{}^\Omega$ |
| $1/2$ | | ${}_P{}^D, {}_\Omega{}^P$ | $P\Omega$ | ${}_{DD}{}^D, {}_{DP}{}^P, {}_{D\Omega}{}^\Omega, {}_{PP}{}^\Omega$ |
| $1$ | | ${}_\Omega{}^D$ | $\Omega\Omega$ | ${}_{DP}{}^D, {}_{D\Omega}{}^P, {}_{PP}{}^P, {}_{P\Omega}{}^\Omega$ |

where $S$ are extra, nonderivative, local "Lorentz" and scale transformations, and $\mathscr{h}_{\mathscr{a}}{}^{\mathscr{b}}$ are the linearized $\mathcal{E}_{\mathscr{a}}{}^{m}$. (Without the currents $S$ and $\Sigma$, only the torsions continue, until the dimension 5/2 $T_{\Omega\Omega}{}^{D}$.) The metric ($G$) constraints ("orthogonality") apart from $G_{PP}{}^{\mathscr{a}}$& $G_{D\Omega}{}^{\mathscr{a}}$, are included in torsion constraints now. They are easier to solve than the full torsion ($T$) constraints because they are purely algebraic, and should be solved first at any particular dimension.

## 3.1 Dimension −1/2

We use $\Lambda^{D}$ and $T_{DD}{}^{\Omega} = 0$ to algebraically fix $H_{D}{}^{P}$ and $\Lambda^{P}$ as $D$ derivatives of $H_{D}{}^{\Omega}$ and $\Lambda^{\Omega}$.

$$\begin{aligned} T_{\alpha\beta\gamma}{}^{\mathscr{a}} =& - H_{(\alpha}{}^{a}\gamma^{b}_{\beta)\gamma}\eta_{ab}{}^{\mathscr{a}} - \gamma^{a}_{\alpha\beta}H_{a}{}^{\mathscr{a}}{}_{\gamma} + \nabla_{(\alpha}H_{\beta)\gamma}{}^{\mathscr{a}} + \nabla_{\gamma}H_{\alpha\beta}{}^{\mathscr{a}} \\ =& - H_{(\alpha}{}^{a}\gamma^{b}_{\beta\gamma)}\eta_{ab}{}^{\mathscr{a}} + \nabla_{(\alpha}H_{\beta\gamma)}{}^{\mathscr{a}} \\ &+ H_{\gamma}{}^{a}\gamma^{b}_{\alpha\beta}\eta_{ab}{}^{\mathscr{a}} - \gamma^{a}_{\alpha\beta}H_{a}{}^{\mathscr{a}}{}_{\gamma} \\ &+ \frac{1}{2}(\nabla_{\alpha}H_{[\beta\gamma]}{}^{\mathscr{a}} + \nabla_{\beta}H_{[\alpha\gamma]}{}^{\mathscr{a}}) \end{aligned}$$

The second term and the third term tell us $G_{DP}{}^{\mathscr{a}} = 0$ and $G_{DD}{}^{\mathscr{a}} = 0$ respectively, from which we can completely solve for $H_{P}{}^{\Omega}$ in terms of $H_{D}{}^{P}$ and have $H_{D}{}^{\Omega} = H_{(\alpha,}{}^{\mathscr{b}}{}_{\beta)} = 10 \otimes 5 = 35 \oplus 10 \oplus 5$. Then $\Lambda^{P}$ algebraically gauges away the 10:

$$\gamma^{\alpha\beta}_{a}\eta^{ab}{}_{\mathscr{a}}H_{\alpha\beta}{}^{\mathscr{a}} = 0 \rightarrow \Lambda^{a} = -\gamma^{\alpha\beta}_{b}\eta^{ab}{}_{\mathscr{a}}\nabla_{\alpha}\Lambda_{\beta}{}^{\mathscr{a}}$$

The 5 is the desired piece; the 35 will be dealt with below.

The first term decomposes into the representations

$$T_{(\alpha,\beta,}{}^{\mathscr{c}}{}_{\gamma)} = 20 \otimes 5 = 64 \oplus 20 \oplus 16\ , \quad H_{\alpha}{}^{b} = 10 \otimes 4 = 20 \oplus 16 \oplus 4$$

So the torsion constraint fixes the 20 and 16, while the 4 is gauge fixed by $\Lambda^{D}$. (Because of gauge invariance, it drops out of this torsion.)

The final step for this dimension uses the remaining, highest-spin differential equation part of this torsion constraint (64) to fix the highest-spin part of $H_{D}{}^{\Omega}$ (35) to be pure gauge of the highest-spin part of $\Lambda^{\Omega}$ (16):

$$\Lambda^{\mathscr{a}}_{\alpha} = 5 \otimes 4 = 16 \oplus 4$$

Thus, converting the $\mathscr{a}$ indices on $T_{(\alpha,\beta,}{}^{\mathscr{c}}{}_{\gamma)}, H_{(\alpha,}{}^{\mathscr{b}}{}_{\beta)}, \Lambda^{\mathscr{a}}_{\alpha}$ into spinor pairs $\langle\alpha\beta\rangle$, the highest-spin parts are eliminated as

$$T_{(\alpha\beta\gamma\delta)\varepsilon} \rightarrow H_{(\alpha\beta\gamma)\delta} \rightarrow \Lambda_{(\alpha\beta)\gamma}$$

where the arrows indicate action by a spinor derivative $d_\alpha$ and symmetrization of its index with the already-symmetrized indices, as $T = dH = 0$ and $\delta H = d\Lambda$. (The totally symmetrized tensors vanish by definition. The traces vanish because: $tr\ T$ has already been solved to vanish; $tr\ H$ was gauged away with $\Lambda^P$; $tr\ \Lambda$ is not used for this purpose.)

At this point the only remaining independent part of all the vielbein with dimension $-1$ or $-1/2$ is the 5 of ${H_D}^\Omega$, and the only independent piece of all of $\Lambda^A$ is the 4 of $\Lambda^\Omega$. Explicitly, we have

$$ {H_D}^\Omega = {H_{\alpha\beta}}^{\theta\phi} = {H_{(\alpha}}^{\langle\theta}\delta^{\phi\rangle}_{\beta)} \,, \quad \Lambda^\Omega = \Lambda^{\theta\phi}_\alpha = \delta^{\langle\theta}_\alpha \lambda^{\phi\rangle} $$

The gauge transformation of prepotential is then (with $\Lambda^P$ in terms of $\nabla_D\Lambda^\Omega$) :

$$ \begin{aligned} &\delta {H_{(\alpha}}^{\langle\theta}\delta^{\phi\rangle}_{\beta)} = -\nabla_{(\alpha}\delta^{\langle\theta}_{\beta)}\lambda^{\phi\rangle} - \nabla^{\langle\phi}\delta^{\theta\rangle}_{(\beta}\lambda_{\alpha)} \\ &\rightarrow \delta H_{\langle\alpha\beta\rangle} = d_{\langle\alpha}\lambda_{\beta\rangle} \end{aligned} $$

reproduces the results of [5].

We also have (with P converted into spinor notation):

$$ {H_D}^P = {H_\alpha}^{\beta\gamma} = \nabla^{(\beta}{H_\alpha}^{\gamma)} $$

$$ {H_P}^\Omega = {H_{a,}}^{\mathscr{b}}{}_\beta = {\eta_{ac}}^{\mathscr{b}}{H_\beta}^c $$

## 3.2 Dimension 0

The constraints of dimension $0, 1/2, 1$ are then solved to find the off-shell field strength, whose vanishing on shell gives the field equations. The dimension 0 objects are ${T_{DD}}^P$, ${T_{DP}}^\Omega$, ${G_{PP}}^{\mathscr{a}}$, ${G_{D\Omega}}^{\mathscr{a}}$.

First, solving ${G_{PP}}^{\mathscr{a}}$ and ${G_{D\Omega}}^{\mathscr{a}}$ at linearized level

$$ \begin{aligned} 0 &= {H_{(a}}^m {\eta_{|m|b)}}^{\mathscr{a}} - {\eta_{ab}}^m {\hbar_m}^{\mathscr{a}} \\ 0 &= {H_\alpha}^\beta \delta^{\mathscr{a}}_{\mathscr{b}} + {H_{\mathscr{b}}}^\beta{}_\alpha{}^{\mathscr{a}} - \delta^\beta_\alpha {\hbar_{\mathscr{b}}}^{\mathscr{a}} \end{aligned} $$

gives us:

$$ \begin{aligned} H^{\mathscr{a}\mathscr{b}}{}_{mn} &= -{\hbar_{[m}}^{[\mathscr{a}}\delta^{\mathscr{b}]}_{n]} \\ {H_{\mathscr{a}}}^\beta{}_\alpha{}^m &= \delta^\beta_\alpha {\hbar_m}^{\mathscr{a}} - {H_\alpha}^\beta \delta^m_{\mathscr{a}} \end{aligned} $$

And furthermore, it guarantees the equivalence of two torsion conditions, up to an extra $\mathcal{U}$ term.

$$\begin{aligned}
T_{DP}{}^{\Omega} - f_{DP}{}^{\Omega} &= H_{\alpha}{}^{\gamma}\gamma^{b}_{\gamma\beta}\eta_{ba}{}^{\mathscr{a}} + H_{a}{}^{b}\gamma^{c}_{\alpha\beta}\eta_{cb}{}^{\mathscr{a}} - H_{\beta}{}^{\mathscr{a}\gamma}{}_{\mathscr{b}}\gamma^{b}_{\alpha\gamma}\eta_{ba}{}^{\mathscr{b}} \\
&+ \nabla_{\alpha}H_{a\beta}{}^{\mathscr{a}} - \nabla_{a}H_{\alpha\beta}{}^{\mathscr{a}} + \eta_{ab}{}^{\mathscr{a}}\nabla_{\beta}H_{\alpha}{}^{b} \\
&= \eta_{ba}{}^{\mathscr{a}}(H_{(\alpha}{}^{\gamma}\gamma^{b}_{\beta)\gamma} - \gamma^{c}_{\alpha\beta}H_{c}{}^{b} + \nabla_{(\alpha}H_{\beta)}{}^{b} - \eta^{bc}{}_{\mathscr{b}}\nabla_{c}H_{\alpha\beta}{}^{\mathscr{b}}) - U^{c\mathscr{a}}_{a\mathscr{b}}\nabla_{c}H_{\alpha\beta}{}^{\mathscr{b}} \\
&= \eta_{PP}{}^{\mathscr{a}}(T_{DD}{}^{P} - f_{DD}{}^{P}) + U(H_{D}{}^{\Omega})
\end{aligned}$$

Now the solution of $H_{\Omega}{}^{\Omega}$ may have a different behavior under gauge transformation. Due to the worldvolume vielbein, now its gauge transformation has an extra piece:

$$\delta H_{\mathscr{a}}{}^{\beta}{}_{\alpha}{}^{m} = -\delta^{m}_{\mathscr{a}}\nabla_{\alpha}\Lambda^{\beta} + \delta^{\beta}_{\alpha}U^{bm}_{a\,\mathscr{a}}\nabla_{b}\lambda^{a}$$

Yet this extra term may not be too harmful, since $H_{\Omega}{}^{\Omega}$ would only show up in $\triangleright_{\Omega}$ or $G_{D\Omega}{}^{\mathscr{a}}\mathcal{D}_{\mathscr{a}}$. Noting that we can find a worldvolume differential in both of these two terms (keeping in mind that $\Omega = \partial\theta$), the extra term can be dropped via $\mathcal{U}$ constraint.

Now we move on to the torsion constraint $T_{DD}{}^{P} = f_{DD}{}^{P}$. At linearized level:

$$H_{(\alpha}{}^{\mu}\Gamma^{a}_{|\mu|\beta)} - \gamma^{m}_{\alpha\beta}H_{m}{}^{a} + \nabla_{(\alpha}H_{\beta)}{}^{a} + \eta^{ab}{}_{m}\nabla_{b}H_{\alpha\beta}{}^{m} = 0$$

Decomposing the vielbeins into irreducible representations, we have $H_{D}{}^{D} = 4 \otimes 4 = 5 \oplus 10 \oplus 1$ and $H_{P}{}^{P} = 5 \otimes 5 = 14 \oplus 10 \oplus 1$. Then for the last two terms, we have conceptually a decomposition of $10 \otimes 10 \to 35 \oplus 35' \oplus 14 \oplus 10 \oplus 5 \oplus 1$. Noting that the $10 \oplus 1$ in the vielbeins represents the local Lorentz $\oplus$ scale transformation; we can gauge fix one set of them, say those in $H_{D}{}^{D}$, with the help of $\Lambda^{S}$. Putting in the solution of $H_{D}{}^{P}$, we can see that the totally symmetrized $35'$ ($\square\square\square\square$) does not show up, as the prepotential lives in the antisymmetric 5. The 35 mixed symmetry ($\square\square\square$ over $\square$) components from $\nabla_{D}H_{D}{}^{P}$ and $\nabla_{P}H_{D}{}^{\Omega}$ cancel each other. Therefore the net result of the last two terms now looks like:

$$\begin{aligned}
&\nabla_{(\alpha}\nabla^{(\gamma}H_{\beta)}{}^{\delta)} - \frac{1}{2}\nabla_{(\alpha}{}^{(\gamma}H_{\beta)}{}^{\delta)} + \frac{1}{2}\nabla_{\theta}{}^{(\gamma}\delta^{\delta)}_{(\alpha}H_{\beta)}{}^{\theta} \\
=&\frac{1}{2}\Delta_{(\alpha}{}^{(\gamma}H_{\beta)}{}^{\delta)} \to (14 \oplus 1) \\
-&\frac{1}{4}(\nabla^{\theta(\gamma}\delta^{\delta)}_{(\alpha}H_{\beta)\theta} - \nabla^{\theta}{}_{(\beta}\delta^{(\delta}_{\alpha)}H^{\gamma)}{}_{\theta}) \to (10) \\
+&\frac{1}{4}(\nabla_{\theta}{}^{(\gamma}\delta^{\delta)}_{(\alpha}H_{\beta)}{}^{\theta} + \nabla_{\theta(\beta}\delta^{(\delta}_{\alpha)}H^{\gamma)\theta}) - \frac{1}{8}\delta^{(\gamma}_{(\alpha}\Delta H_{\beta)}{}^{\delta)} \to (5)
\end{aligned}$$

with

$$\Delta_{\alpha}{}^{\gamma} = [D_{\alpha}, D^{\gamma}] - tr\ , \Delta = tr$$

We can read out the solutions:

$$H_P{}^P = H_{\alpha\beta}{}^{\gamma\delta} = \frac{1}{2}\Delta_{(\alpha}{}^{(\gamma}H_{\beta)}{}^{\delta)} - \frac{1}{4}(\nabla^{\theta(\gamma}\delta^{\delta)}_{(\alpha}H_{\beta)\theta} - \nabla^{\theta}{}_{(\beta}\delta^{(\delta}_{\alpha)}H^{\gamma)}{}_{\theta})$$
$$H_D{}^D = (H_D{}^D)_\alpha{}^\beta = \frac{1}{4}(\nabla^{\theta\beta}H_{\alpha\theta} + \nabla^{\theta}{}_{\alpha}H^{\beta}{}_{\theta}) + \frac{1}{8}\Delta H_\alpha{}^\beta$$

## 3.3 Dimension 1/2

Objects with dimension 1/2 are $T_{DD}{}^D$, $T_{DP}{}^P$, $T_{D\Omega}{}^\Omega$ and $T_{PP}{}^\Omega$. We can start with $T_{DD}{}^D$:

$$T_{\alpha\beta}{}^\delta = H_{(\alpha|}{}^S f_{S|\beta)}{}^\delta - \gamma_{\alpha\beta}{}^a H_a{}^\delta + \nabla_{(\alpha}(H_D{}^D)_{\beta)}{}^\delta$$

which contains $20 \oplus 16 \oplus 4$ in total. Picking up the Lorentz covariant $16 \oplus 4$, we have:

$$(H_P{}^D)_{(16\oplus4)} = (H_{\alpha\beta}{}^\delta)_{(16\oplus4)} = \nabla_{(\alpha}(H_D{}^D)_{\beta)}{}^\delta$$

Then for $T_{DP}{}^P$, we have:

$$T_{DP}{}^P = T_{\alpha a}{}^b = H_\alpha{}^S f_{Sa}{}^b - H_a{}^\beta\gamma^b_{\alpha\beta} + \gamma^c_{\alpha\beta}\eta_{ca}{}^{\mathscr{a}}H_{\mathscr{a}}{}^{\beta b} + \nabla_\alpha H_a{}^b - (\delta^d_a\delta^b_c - \eta_{ac}{}^{\mathscr{a}}\eta^{bd}{}_{\mathscr{a}})\nabla_d H_\alpha{}^c$$
$$= 64 \oplus 40 \oplus (20\times 2) \oplus (16\times 3) \oplus (4\times 2)\ (naively)$$

In front of $\nabla_P H_D{}^P$, we have a projector, which, in spinor indices ($a \to (a_1a_2)$), looks like (switching b,c with Sp(4) metric):

$$\delta^d_a\delta^c_b - \eta_a{}^{c\mathscr{a}}\eta^d{}_{b\mathscr{a}}$$
$$=\delta^{(c_1}_{(a_1}\delta^{c_2)}_{a_2)}\delta^{(d_1}_{(b_1}\delta^{d_2)}_{b_2)} + \delta^{(d_1}_{(a_1}\delta^{d_2)}_{a_2)}\delta^{(c_1}_{(b_1}\delta^{c_2)}_{b_2)} - \frac{1}{2}\delta^{(c_1|}_{(a_1}\delta^{(d_1}_{a_2)}\delta^{d_2)}_{(b_1}\delta^{|c_2)}_{b_2)} - \frac{1}{2}\delta^{(c_1}_{(a_1}C_{a_2)(b_1}C^{c_2)(d_1}\delta^{d_2)}_{b_2)}$$

Then masked by the projector, the 64 (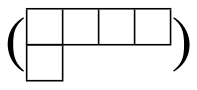) would be filtered out:

$$(\delta^d_a\delta^c_b - \eta_a{}^{c\mathscr{a}}\eta^d{}_{b\mathscr{a}})\nabla_d H_{\alpha c}$$
$$=(\nabla_{a_1a_2}\nabla_{(b_1}H_{|\alpha|b_2)} + \nabla_{b_1b_2}\nabla_{(a_1}H_{|\alpha|a_2)} - \frac{1}{2}\nabla_{(b_1|(a_1}\nabla_{a_2)}H_{\alpha|b_2)} - \frac{1}{2}\nabla_{(a_1|(b_1}\nabla_{b_2)}H_{\alpha|a_2)})$$
$$-\frac{1}{2}(\nabla_{(b_1}{}^\beta C_{b_2)(a_1|}\nabla_\beta H_{\alpha|a_2)} + \nabla_{(b_1}{}^\beta C_{b_2)(a_1}\nabla_{a_2)}H_{\alpha\beta})$$
$$=\frac{1}{4}\delta^{\langle\mu|}_{(a_1}\delta^{\langle\gamma}_{a_2)}\delta^{\phi\rangle}_{(b_1}\delta^{|\nu\rangle}_{b_2)}\nabla_{\mu\gamma}\nabla_\phi H_{\alpha\nu} \to (40\oplus16\oplus4)$$
$$+\frac{1}{4}\delta^{(\gamma}_{(a_1}C_{a_2)(b_1}\delta^{\phi)}_{b_2)}\nabla_\gamma{}^\beta\nabla_\phi H_{\alpha\beta} \to (20\oplus16\oplus4)$$
$$+\frac{1}{4}\delta^{(\gamma}_{(a_1}C_{a_2)(b_1}\delta^{\phi)}_{b_2)}\nabla_\gamma{}^\beta\nabla_\beta H_{\alpha\phi} - \frac{1}{8}C_{(a_1|(b_1}C_{|a_2)|b_2)}\nabla^{\theta\beta}\nabla_\beta H_{\alpha\theta} \to (16\oplus4)$$

Combining the first two terms with the $4 \otimes (14 \oplus 1)$ and $4 \otimes 10$ of $\nabla_D H_P{}^P$ respectively would cancel the 40 (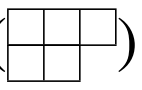) and 20 (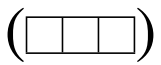) and give us:

$$
\begin{aligned}
&\frac{1}{4}\delta^{\langle\mu|}_{(a_1}\delta^{\langle\gamma}_{a_2)}\delta^{\phi\rangle}_{(b_1}\delta^{|\nu\rangle}_{b_2)}(\frac{1}{2}\nabla_\alpha \Delta_{\gamma\phi}H_{\mu\nu} - \nabla_{\mu\gamma}\nabla_\phi H_{\alpha\nu}) \\
=&\frac{1}{4}\delta^{\langle\mu|}_{(a_1}\delta^{\langle\gamma}_{a_2)}\delta^{\phi\rangle}_{(b_1}\delta^{|\nu\rangle}_{b_2)}(\frac{1}{2}\nabla_{\alpha\gamma}\nabla_\phi H_{\mu\nu} - \nabla_{\mu\gamma}\nabla_\phi H_{\alpha\nu} - \epsilon_{\alpha\gamma\phi\theta}(\nabla^3)^\theta H_{\mu\nu}) \to (16 \oplus 4)
\end{aligned}
$$

where the $(\nabla^3)^\theta$ stands for $\frac{1}{6}\epsilon^{\alpha\beta\gamma\theta}\nabla_\alpha\nabla_\beta\nabla_\gamma$, and

$$
\frac{1}{4}\delta^{(\gamma}_{(a_1}C_{a_2)(b_1}\delta^{\phi)}_{b_2)}\nabla^\theta{}_\gamma\nabla_{[\alpha}H_{\phi]\theta} \to (16 \oplus 4)
$$

Based on the symmetry property of spacetime momenta indices, we can then identify these two sets of $16 \oplus 4$ with $H_\Omega{}^P$ and $H_D{}^S$.

### 3.4 Dimension 1

We can take a look into $T_{PD}{}^D$:

$$
T_{a\alpha}{}^\beta = H_a{}^S f_{S\alpha}{}^\beta - \gamma^b_{\alpha\delta}\eta_{ba}{}^{\mathcal{a}} H_{\mathcal{a}}{}^{\delta\beta} + \nabla_a (H_D{}^D)_\alpha{}^\beta - \nabla_\alpha H_a{}^\beta
$$

whose Lorentz covariant part, after putting in the solution of $H_P{}^D$, can help us finding a solution of $H_\Omega{}^D$:

$$
\begin{aligned}
&\nabla_{(\gamma}\nabla_{\delta)}(H_D{}^D)_{\alpha\beta} - \frac{1}{2}\nabla_{\langle\alpha}\nabla_{(\delta}(H_D{}^D)_{\gamma)\beta\rangle} \\
=&\frac{1}{2}\delta^{(\mu|}_{(\gamma}\delta^{\langle\nu}_{\delta)}\delta^{\theta\rangle}_{\langle\alpha}\delta^{|\phi)}_{\beta\rangle}\nabla_\mu\nabla_\nu(H_D{}^D)_{\theta\phi}
\end{aligned}
$$

Conceptually we have $35 \oplus 10 \oplus 5$, yet only the 5 remains. There are two 5 representations in $H_\Omega{}^D = 35 \oplus 14 \oplus 10 \oplus (5 \otimes 2) \oplus 1$, and the form of the projector determines that the structure should be:

$$
H_\Omega{}^D = H^{\alpha\beta}_{\mathcal{a}} = A^{\mathcal{b}}\gamma^{\alpha\beta}_{\mathcal{b}\mathcal{a}}
$$

# 4 Conclusions

Using current algebra, we can construct the 3-dimensional F-theory gravitational supermultiplet directly from solving a set of torsion constraint. We list the solutions here:

| | | | |
|---|---|---|---|
| ${H_D}^{\Omega}$ | $H_{\alpha\beta}$ | 5 | prepotential |
| ${H_D}^{P}, {H_P}^{\Omega}$ | $\nabla_{(\alpha} H_{\vert\beta\vert\gamma)}$ | 16⊕4 | |
| ${H_P}^{P}$ | $\frac{1}{8}\delta^{\langle\mu\vert}_{(\alpha}\delta^{\langle\rho}_{\beta)}\delta^{\sigma\rangle\vert}_{(\delta}\delta^{\nu\rangle}_{\gamma)}\Delta_{\mu\nu}H_{\rho\sigma} + \frac{1}{4}\delta^{(\mu}_{(\alpha}C_{\beta)(\delta}\delta^{\nu)}_{\gamma)}\nabla^{\theta}{}_{\mu}H_{\nu\theta}$ | 14⊕10⊕1 | graviton |
| ${H_D}^{D}$ | $\frac{1}{4}\nabla^{\theta}{}_{\langle\beta}H_{\alpha\rangle\theta} + \frac{1}{8}\Delta H_{\alpha\beta}$ | 5 | |
| ${H_P}^{D}$ | $\nabla_{(\alpha}({H_D}^{D})_{\beta)\delta}$ | 16⊕4 | gravitino |
| ${H_\Omega}^{P}$ | $\frac{1}{4}\delta^{\langle\mu\vert}_{(a_1}\delta^{\langle\gamma}_{a_2)}\delta^{\phi\rangle}_{(b_1}\delta^{\vert\nu\rangle}_{b_2)}(\frac{1}{2}\nabla_{\gamma[\alpha}\nabla_{\vert\phi\vert}H_{\mu\nu]} - \epsilon_{\alpha\gamma\phi\theta}(\nabla^3)^{\theta}H_{\mu\nu})$ | 16⊕4 | |
| ${H_\Omega}^{D}$ | $\frac{1}{2}\delta^{(\mu\vert}_{(\gamma}\delta^{\langle\nu}_{\delta)}\delta^{\theta\rangle}_{\langle\alpha}\delta^{\vert\phi)}_{\beta\rangle}\nabla_{\mu}\nabla_{\nu}({H_D}^{D})_{\theta\phi}$ | 5 | auxiliary |

We can see that these components are in full consistency with the previous results.

The net form of gauge transformation of the gauge parameter also shows up automatically. As the 16 of $\Lambda^{\Omega}$ is used to gauge fix the 35 of ${H_D}^{\Omega}$, we need to constraint the 16 part of $\delta\Lambda^{\Omega}$ being 0 up to a $\mathcal{U}$ term when doing worldvolume diffeomorphism. The $\mathcal{U}$ term itself can be gauge fixed by another gauge transformation of $\Lambda^{\Omega}$.

A generalization to higher dimension would be straightforward.

# Acknowledgements

This work was supported by NSF grant PHY-1620628.

# A Indices

| space | representation | flat | curved |
|---|---|---|---|
| worldvolume | vector | $\mathscr{a}$ | $\mathscr{m}$ |
| superspace | super | $A$ | $M$ |
| | $D$ (spinor) | $\alpha$ | $\mu$ |
| | $P$ (vector) | $a$ | $m$ |
| | $\Omega$ | ${}_{\mathscr{a}}{}^{\alpha}$ | ${}_{\mathscr{m}}{}^{\mu}$ |
| gauge | gauge(5 in 3D case) | $\varpi$ | |

$$(AB] = AB + BA\ , \quad [AB) = AB - BA$$

All superindices (e.g., in the above equations) have implicit relative $-$ signs from pushing fermionic (spinor) index parts past fermionic. But ( ) and [ ] have $+$ and $-$, respectively, for any non-superindices. For supertraces

$${}^{A}{}_{A} = {}^{a}{}_{a} + {}^{\alpha}{}_{\alpha} - {}^{\mathscr{a}}{}_{\alpha\mathscr{a}}{}^{\alpha}$$

when adjacent with that ordering.

The symplectic-trace-free part of [ ] is $\langle\ \rangle$, and the orthogonal-trace-free part of ( ) is { } (not used in this paper).

For abbreviation and identification, we sometimes replace parts of superindices with the symbols for their currents $D, P, \Omega$.